\documentclass[%
 reprint,
 amsmath,amssymb,
 aps,
]{revtex4-2}

\usepackage{graphicx}
\usepackage{bm}
\usepackage{graphicx}
\usepackage{dcolumn}
\usepackage{bm}
\usepackage{color}
\usepackage{CJK}
\usepackage{fancyhdr}
\usepackage{calc}
\usepackage{lineno}
\usepackage{float}
\usepackage{mpgmpar} 
\usepackage{amsmath,amsthm} 
\usepackage{bm,here}
\usepackage{lipsum}
\usepackage{booktabs}

\begin{document}

\preprint{APS/123-QED}
\title{Dynamic tuning of ENZ wavelength in conductive polymer films via polaron excitation}
\author{Hongqi Liu$^{1}$, Junjun Jia$^{2*}$, Menghui Jia$^{3}$, Chengcan Han$^{1}$, Sanjun Zhang$^{3}$, Hui Ye$^{1*}$, Heping Zeng$^{3}$}
\email{jia@aoni.waseda.jp; huiye@zju.edu.cn; hpzeng@phy.ecnu.edu.cn.}
\affiliation{[1] College of Optical Science and Engineering, Zhejiang University, Hangzhou 310027, China.\\
[2] Global Center for Science and Engineering (GCSE), Faculty of Science and Engineering, Waseda University, Tokyo 169--8555, Japan.\\
[3] State Key Laboratory of Precision Spectroscopy, East China Normal University, Shanghai, 200062, China.}

\date{\today}

\begin{abstract}
Traditional metal and $n$--type doped semiconductor materials serve as emerging epsilon--near--zero (ENZ) materials, showcasing great potential for nonlinear photonic applications. However, a significant limitation for such materials is the lack of versatile ENZ wavelength tuning, and thus dynamic tuning of the ENZ wavelength remains a technical challenge, thereby restricting their potential applications, such as multi--band communications. Here, dynamic tuning of the ENZ wavelength in $p$--type organic PEDOT: PSS films is achieved through a reversible change in hole concentrations originated from the polaron formation/decoupling following optical excitation, and a tunable ENZ wavelength shift up to 150 nm is observed. Experimental investigations about ultrafast dynamics of polaron excitation reveal a $\sim$80 fs time constant for polaron buildup and a $\sim$280 fs time constant for polaron decoupling, indicating the potential of reversal ultrafast switching for the ENZ wavelength within subpicosecond time scale. These findings suggest that $p$--type organic semiconductors can serve as a novel platform for dynamically tuning the ENZ wavelength through polaron excitation, opening new possibilities for ENZ--based nonlinear optical applications in flexible optoelectronics. 
\end{abstract}

\maketitle

\section{Introduction}
Epsilon--near--zero (ENZ) materials exhibit a real part of the permittivity ($\epsilon$) crossing zero within the infrared to visible spectral region, accompanied by a sufficiently small imaginary part, leading to a vanishingly small refractive index, also known as near--zero--index (NZI) conditions \cite{Fomra2024, Reshef2019}. These unique properties give rise to peculiar optical phenomena, such as enhanced electric file, slow light, wavelength expansion, alongside diverging group and phase velocities \cite{Kinsey2019}. 

The ENZ wavelength (or frequency) of a material depends on its carrier concentration. Metals, with carrier concentrations as high as 10$^{23}$ cm$^{-3}$, exhibit ENZ wavelengths in the visible region. In contrast, degenerate semiconductors doped with impurity atoms, such as transparent conductive oxides (TCO), have carrier concentrations up to 10$^{22}$ cm$^{-3}$, and their ENZ wavelengths are located in the near--infrared region \cite{Zhao2015, Jia2024}. These materials have emerged as promising platforms for ENZ--based optical applications, including ultrafast all--optical modulation \cite{Kinsey2015}, terahertz generation \cite{Guo2017, Jia2021}, nonlinear nano-optics \cite{Alam2016}, giant enhancements of electric fields \cite{Campione2013}, and harmonic generation \cite{Capretti2015, Yang2019}. However, the ENZ wavelength is typically fixed after fabrication, as altering the carrier density of traditional ENZ materials is highly challenging. This limitation restricts the post--fabrication tunability of the ENZ wavelength. Furthermore, integrating these inorganic materials into flexible electronic devices presents additional challenges. Therefore, discovering a new material system with tunable ENZ wavelength is of critical importance.

To date, there has been limited research on ENZ materials with tunable wavelength. In metals or doped semiconductors, ultrafast electron gas heating has been shown to induce a shift in ENZ wavelength on sub--100 fs timescales \cite{Caspani2016, Alam2016, Fatti2000}, resulting in frequency changes of up to a few percent of the carrier frequency \cite{Zhou2020, Wang2019, Yang2017}. However, due to intrinsic excitation threshold, relatively high optical losses, and short interaction durations, most reported nonlinear ENZ frequency shifts require high--intensity laser sources, with intensities reaching several hundreds of gigawatts per squared centimeter \cite{LiuC2021, Bruno2020, Bohn2021}, thereby limiting broader applications of traditional ENZ materials. Therefore, developing a method to achieve significant ENZ wavelength shifts under lower laser intensities is essential for expanding their utility.

\begin{figure*}[t]
\begin{center}
\includegraphics[clip, width=18cm]{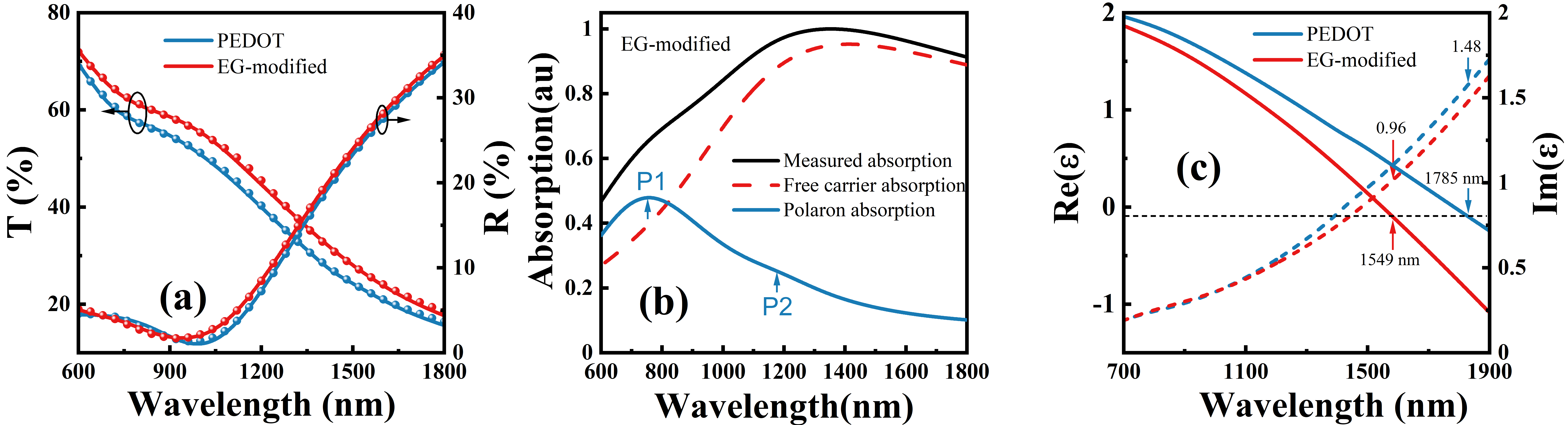}
\caption{(a): Transmittance and reflectance spectra of pristine (blue line) and EG--modified (red line) PEDOT films. The measured spectra (dotted line) were fitted well by the Drude--Lorentz model (solid lines). (b): Measured absorption spectra for the EG--modified film, which are well reproduced by the Drude--Lorentz model with the consideration of free carrier (dashed line) and polaron (blue solid line) absorptions, revealing distinct polaronic features with two peaks at 752 nm (P1) and 1190 nm (P2). (c): Permittivity of the pristine and EG--modified PEDOT films.}
\label{Figure1}
\end{center}
\end{figure*}

On the other hand, our previous studies \cite{Hu2023} have demonstrated that organic conductive conjugated polymers can serve as a promising platform for developing novel ENZ applications, featuring by their chemical flexibility and diverse synthetic strategies \cite{Ahsin2024}. These polymers possess carrier concentrations ranging from 10$^{18}$ to $^{22}$ cm$^{-3}$ \cite{Wang2018, Sun2015, Bredas1982}, which can be tuned through the addition of polar solvents and further optimized by engineered crystallization or solvent treatment \cite{Kim2012, Wang2018}. One key physical characteristic of such polymers is the formation of polarons and bipolarons, which represent singly and doubly charged quasiparticals, respectively, that are localized along the polymer backbone due to a strong electron--phonon coupling \cite{Zozoulenko2019}. Recent studies have shown that these polaronic states can be modulated over a wide range through electrical tuning of the polymer's redox state \cite{Chen2020, Karki2022}. Notably, polaronic charge carrier can be tuned to concentrations as high as 2.6 $\times$ 10$^{21}$ cm$^{-3}$ at low voltages, opening an exciting research avenue where tuning polaron states can effectively alter the carrier density and, consequently, the ENZ wavelength. Given the physical nature of polarons and bipolarons, these polaronic states should also be tailored by optical excitations \cite{Kim2008, Kalagi2016}, suggesting the possibility of achieving an ENZ wavelength shift by exciting a significant population of polaronic carriers. However, while most studies have focused on the carrier transport properties of these polymers \cite{Zozoulenko2019, Bubnova2014}, the potential for optically induced ENZ wavelength shifts has yet to be explored.

In this study, we successfully modulated the ENZ wavelength of a $p$--type conjugated polymer film, Poly(3,4-ethylenedioxythiophene) (PEDOT), by tuning its hole densities through polaron excitation. A significant ENZ wavelength shift of up to $\sim$150 nm was achieved in PEDOT films modified by ethylene glycol (EG--modified) under the laser intensities down to several GW/cm$^2$. The underlying mechanisms driving this wavelength shift are discussed in detail. We believe that these findings open new avenues for dynamic tuning of ENZ wavelength, paving the way for advancements in ENZ--based optical applications, particularly in flexible optoelectronics.

\section{Results and Discussion}
\subsection{Optical properties of $p$--type PEDOT films}
The pristine and EG--modified PEDOT films used in this study were spin--coated on the glass substrates (See Methods). Hall measurements revealed that EG modification increases the carrier density ($n$) of PEDOT films from 9.68 $\times$ 10$^{20}$ to 2.16 $\times$ 10$^{21}$ cm$^{-3}$, accompanied by an increase in mobility  ($\mu$) from 0.036 to 2.26 cm$^2$/Vs. Both films exhibit $p$--type conductivity ($\sigma$). The increase in carrier density is attributed to the addition of ethylene glycol, a polar solvent that facilitates the separation of uncoupled PSS from PEDOT, thereby increasing carrier density \cite{Kim2012}. On the other hand, the polar solvent is considered to strengthen $\pi$-$\pi$ coupling among polymer chains, promoting electron delocalization \cite{Lee1993} and resulting in improved mobility.

The ENZ wavelengths of both films were determined from their transmittance and reflectance spectra. Fig. 1(a) shows that both films exhibit a reflectance of approximately 35\% at 1800 nm, which decreases monotonically toward shorter wavelengths, reaching a minimum around 1 $\mu$m. This minimum in reflectance is indicative of metal--like behavior, which can be described by the Drude model with a plasmon frequency ($\omega_p$) \cite{Chen2019}. The EG--modified PEDOT film exhibits a smaller reflectance--minima compared with the pristine PEDOT film, suggesting a higher $\omega_p$ consistent with the increased carrier density in the EG--modified PEDOT film. 

Fig. 1(b) presents the typical absorption spectra of EG--modified PEDOT film. A broad absorption peak centered around 1300 nm is primarily associated with free carrier absorption, while a small shoulder near 750 nm is attributed to polaronic absorption \cite{Zozoulenko2019, Karki2022}. Moreover, a detailed comparison with the pristine film reveals the presence of an additional polaronic band around 1200 nm in the EG--modified PEDOT film. Considering an optical band gap of 3.85 eV, these polaron states (denoted as P1 and P2 in Fig. 1(b)) are identified as in--gap states. The absorption coefficient due to polaron formation ($\alpha_{pol}$) can be expressed as \cite{Emin1993}:
\begin{equation}
\alpha_{pol} = A\exp \left[-\left(2 E_{b}-\hbar \omega\right)^{2} / \Delta^{2}\right]
\end{equation}

where $A$ is a fitting parameter, 2$E_b$ is the energy associated with polaron formation, and $\Delta$ is related to the energy broadening of polaron sites. By fitting the absorption spectra, 2$E_b$ for the polaron P1 and P2 states are estimated to be 1.65 eV and 1.04 eV, respectively.  

\begin{figure}[h]
\begin{center}
\includegraphics[clip, width=8.5cm]{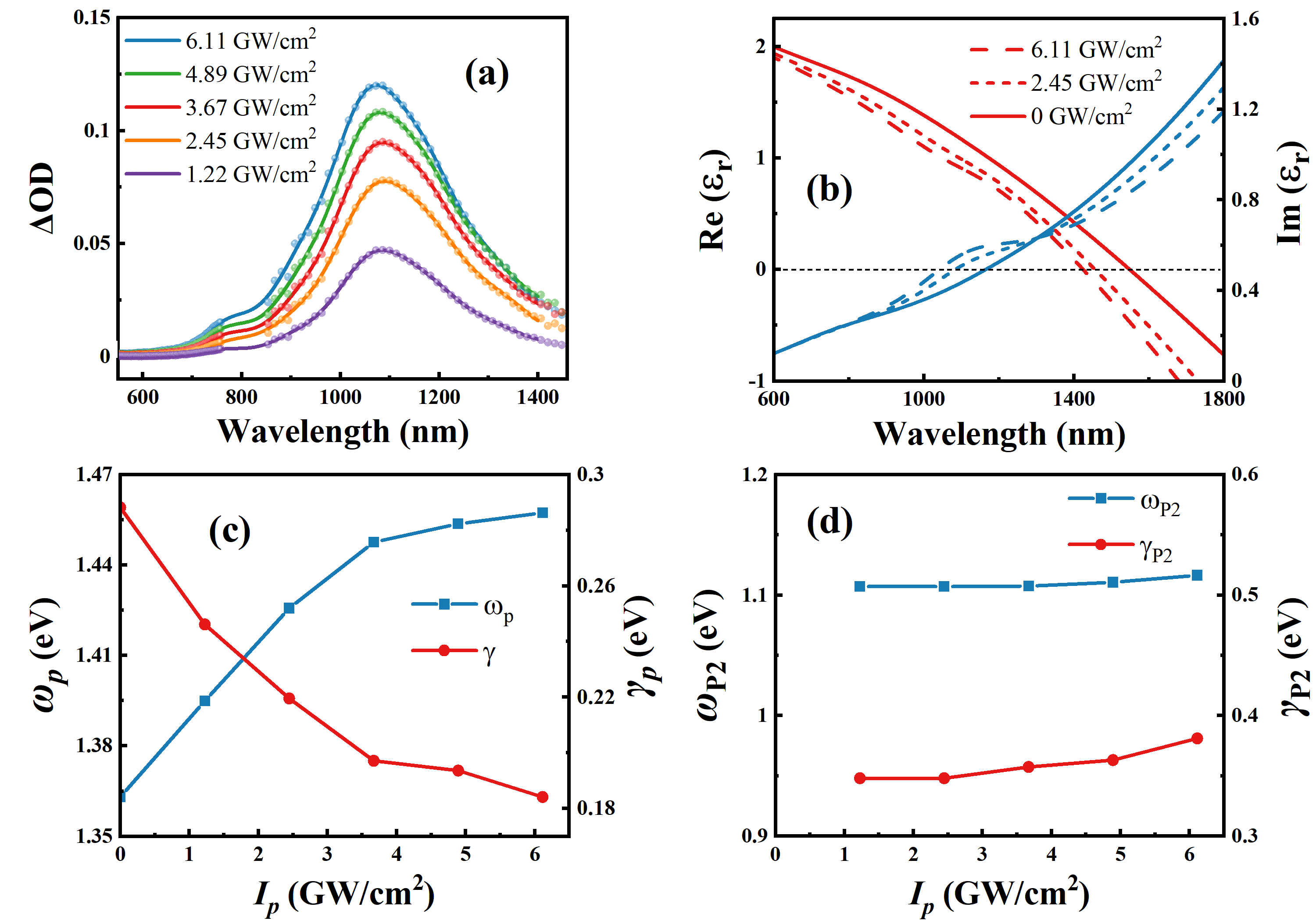}
\caption{(a) Optical density of EG--modified PEDOT film under varying excitation intensities. (b) The fitted permittivity at different excitation intensities, where the zero--crossing permittivity in the real part corresponds to the ENZ wavelength. (c) The fitted plasma frequency increases with excitation intensity. (d) The fitting parameters for the Lorentz oscillator modeling the P2 polaronic state around 1080 nm. Note that the laser excitation wavelength in the experiments is 800 nm.}
\label{Figure2}
\end{center}
\end{figure}

Taking both polaron and free carrier absorptions into consideration, the absorption spectra of PEDOT films can be fitted by the following Lorentz--Drude model:

\begin{equation}		
\varepsilon_r(\omega)= \varepsilon_{\infty}-\frac{\omega_p^2}{\omega(\omega+i\gamma_p)} - \sum\limits_{j=1}^{2}\frac{f_j\omega_j^2}{\omega_j^2 + i\omega\gamma_j - \omega^2}			
\end{equation}

where $\epsilon_\infty$ is the background dielectric constant at high frequency. The second term with the plasmon frequency $\omega_{p}^{2}=\frac{n_he^{2}}{m_h^{*}\varepsilon_{0}}$ and the damping coefficient $\gamma_p=\frac{e}{m_h^*\mu}$ represents free carrier absorption, where $e$ is the elementary charge, $\varepsilon_0$ is the vacuum dielectric constant, $n_h$ is the hole density, and $m_h^*$ is hole effective mass. The third term describes two polaronic contributions, where $f_j$, $\omega_j$, $\gamma_j$ are the amplitude, frequency, broadening of the polaronic Lorentz oscillators, respectively \cite{Emin1993}. Fig. 1(c) shows the fitted complex dielectric constants, showing zero--crossings of the real part of permittivity at 1785 nm for the pristine PEDOT film and at 1549 nm for the EG--modified PEDOT film. Both films exhibit large optical losses (Im($\epsilon$)$\gtrsim$1) at their ENZ wavelengths, larger than that of ITO (Im($\epsilon$)=0.35 at $\lambda_{\texttt{ENZ}}$=1400 nm) \cite{Jia2021}. In comparision, the EG--modified PEDOT film exhibits a lower optical loss at its ENZ wavelength, as indicated by its imaginary part (Im($\epsilon$)=0.94), which can be attributed to its higher mobility. 

\subsection{Tuning ENZ wavelength via polaron excitation}

Dynamic tuning of the ENZ wavelength has been investigated through ultrafast electron gas heating observed on the timescales of 0.1 to 1 ps \cite{Alam2016, Fomra2024}. Due to the $p$--type nature and large band gap (3.85 eV) of PEDOT films \cite{Hu2023}, achieving the same ultrafast electron gas heating in this material system is impossible. Alternatively, we developed a new method for dynamically tuning the ENZ wavelength in PEDOT films, namely through polaron excitation. 

Fig. 2(a) shows the optical densities ($\Delta$OD) of PEDOT films under laser irradiation with varying intensities. A main peak appears at 1100 nm with a shoulder at 745 nm. The main peak is attributed to the P1 polaronic state, while the shoulder corresponds to an in--gap state arising from the formation of the P2 polaronic state. The significant increase in the P1 peak intensity at 1100 nm with laser intensity is attributed to a rise in polaron density. This increase, in principle, introduces more holes in the PEDOT films, which is considered to shift the ENZ wavelength. To understand the effect of polaron excitation on the ENZ wavelength shift, we fitted the permittivity of PEDOT films under varying optical excitations using the Lorentz--Drude model described in Eq. (2). Fig. 2(b) displays the real and imaginary parts of permittivity at different excitation intensities. With the excitation intensity increasing from 0 to 2.12 mJ/cm$^2$, the zero--crossing ENZ wavelength shifts from 1549 nm to 1425 nm, demonstrating the ENZ wavelength shift induced by polaron excitation. Excitingly, the imaginary part at the zero--crossing ENZ wavelength decreases significantly with increasing excitation intensity, as shown in Table \ref{ENZ}. 

\begin{table}[hb]
\caption{Zero--crossing ENZ wavelength and its imaginary part of dielectric constant under varied irradiation intensities.} 
\label{ENZ}
\begin{tabular}{lllllll}
\hline \hline
Power density (GW/cm$^2$)      &  0.00   &  1.22 & 2.45 & 3.67  & 4.90 & 6.11   \\
\hline
ENZ (nm)              &  1549  & 1495  & 1457 & 1431  & 1427 & 1425 \\
Im($\epsilon$)        &  0.96  & 0.85  & 0.78  & 0.72 & 0.72 & 0.71 \\
\hline \hline
\end{tabular} 
\end{table}

Further investigation into the saturation of polaron exciatation is shown in Fig. 2(c). An indicative of ENZ wavelength, $\omega_{p}$, shows an exponential increase toward saturation with increasing excitation intensity. At the same time, the damping parameter $\gamma_p$ in the Drude model shows a decreasing trend, which can be attributed to the increase in hole mobility. Through the following relationship among $\omega_{\mathrm{ENZ}}$, $\omega_p$, and $\gamma_p$,  
    
\begin{equation}		
\omega_{\mathrm{ENZ}}=\sqrt{\frac{\omega_{p}^{2}}{\varepsilon_{\infty}}-\gamma_p^{2}}=\sqrt{\frac{n e^{2}}{\varepsilon_{0} \varepsilon_{\infty} m^{*}}-\gamma_p^{2}} \propto \sqrt{n},   			
\end{equation}
the ENZ wavelength is directly associated to the carrier density. An increase in $\omega_{\mathrm{ENZ}}$ with excitation intensity in Fig. 3(c) indicates a rise in carrier density, specifically due to an increase in hole concentration attributed to polaron formation via optical transitions from the valence band (VB) to in--gap states. This confirms the feasibility of dynamic tuning of ENZ wavelength through such polaron excitation. It is also worthy to note that a minor absorption peak appears around the P2 polaronic state in Fig. 2(a), which is quantitatively described by $\omega_2$ and $\gamma_2$ as shown in Fig. 2(d). The approximately constant $\omega_2$ and $\gamma_2$ suggest that the P2 polaronic state remains unaffected by the laser irradiation of 800 nm.

\subsection{Ultrafast dynamics of polaron excitation}

Since the duration of dynamic tuning is critical for controlling the ENZ wavelength, the pump--probe based femtosecond time--resolved transient absorption measurements were performed to investigate the ultrafast dynamics of polaron excitation. Fig. 3(a) presents ultrafast polaron dynamics in the EG-modified PEDOT film, irradiated by a 1.5 eV pump laser at an intensity of 2.12 mJ/cm$^2$. A distinct peak appears at approximately 1100 nm, corresponding to the P2 polaron state, and gradually diminishes over time. Although a weak shoulder located at the position ($\sim$790 nm) of the P1 polaronic state can also be observed, the P2 polaron state at $\sim$1100 nm) dominates the transient optical absorption. Based on the measurement principle of pump--probe transient absorption spectroscopy, the underlying physics of this transient process can be described as follows: after photoexcitation, the VB electrons are pumped to the P2 polaron states, which are located above the VB maximum, leaving holes in the VB, as depicted in Fig. 3(b), and the electrons trapped in the polaron states are detected by the probe laser due to the Pauli blocking \cite{Jia2021}. Note that these trapped electrons interact with the lattice, gradually relaxing to the initial state. 

\begin{figure}[h]
\begin{center}
\includegraphics[clip, width=8.5cm]{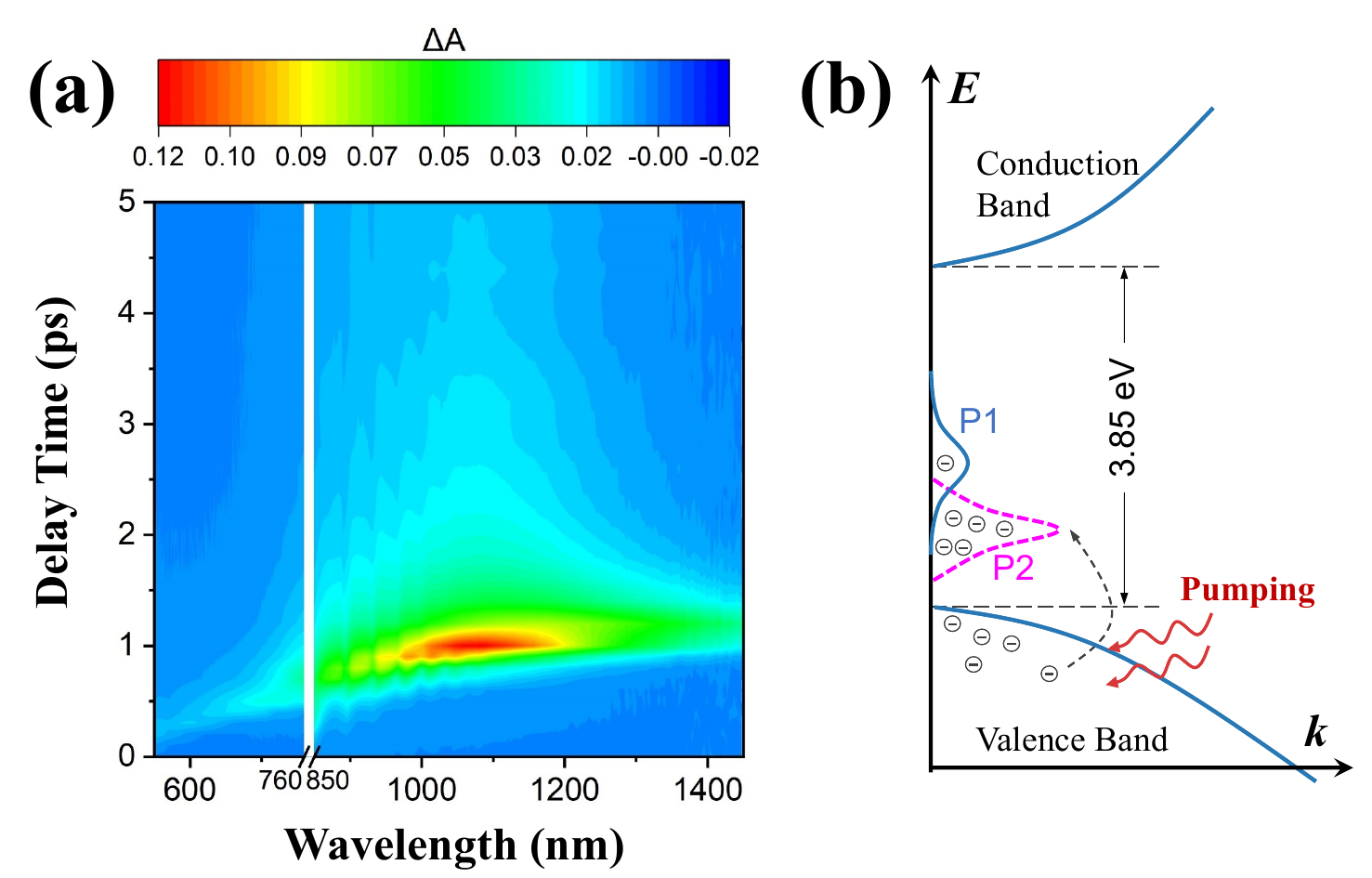}
\caption{(a) Time-resolved transient absorption spectra from the pump--probe measurement. (b) A band diagram with polaron states to explain the buildup and decoupling of polaron states. The measured band gap is 3.85 eV by UV absorption spectroscopy, and the P1 and P2 polaron states are located above valence band maximum.}
\label{Figure3}
\end{center}
\end{figure}

To understand the relationship between polaron absorption and the ENZ wavelength, the transient absorption spectra in Fig. 3(a) were integrated into the static absorption spectra, and subsequently fitted using the Lorentz--Drude model (Eq. 2) to determine the ENZ wavelength. Fig. 4(a) presents the integrated absorption spectra, which are deconvoluted into polaronic absorption and free carrier absorption. A significant polaron absorption peak at the P2 position under laser irradiation is evident. Fig. 4(b) displays the fit plasma frequency and ENZ wavelength as a function of temporal evolution after excitation. A clear blue shift in the ENZ wavelength in the initial time is observed, driven by an increase in hole density ($\omega_p$), highlighting the potential for dynamic tuning of the ENZ wavelength. Such ENZ wavelength shift corresponds to the temporal evolution of polaron density, as shown in Fig. 4(c).  

\begin{figure*}[t]
\begin{center}
\includegraphics[clip, width=18cm]{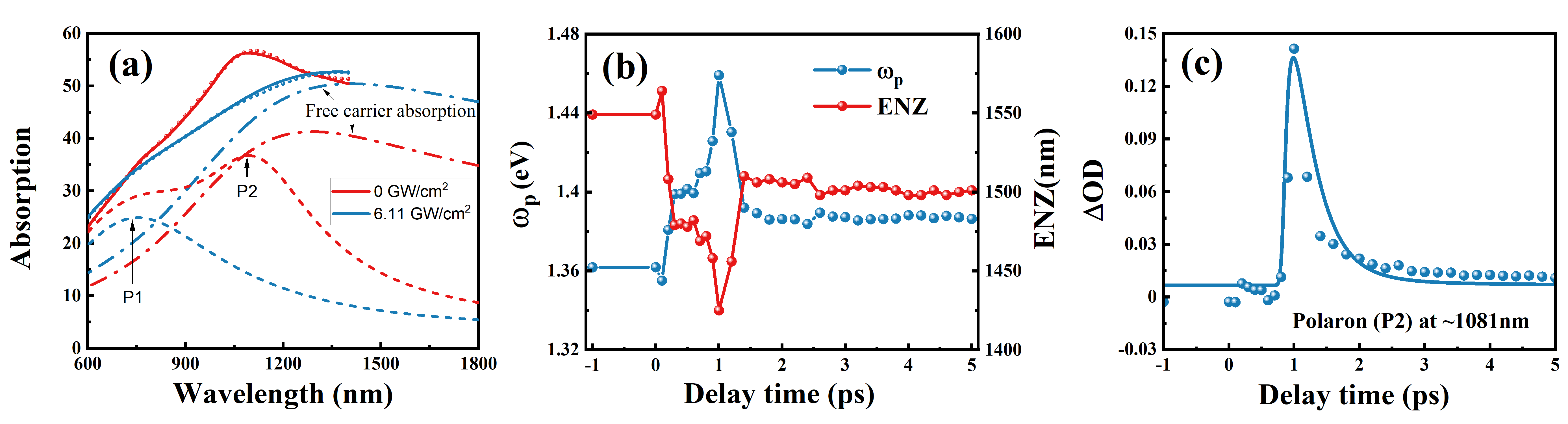}
\caption{(a): The static absorption spectra (0 GW/cm$^2$, scattered points), and the integrated transient absorption spectra with the static absorption spectra (scattered points) under laser irradiation of 6.11 GW/cm$^2$ for the EG–-modified film. Both are fitted using the Lorentz--Drude model with the polaron absorption (dashed lines) and free carrier absorption (dash--dotted lines). (b): The temporal evolution of plasma frequency ($\omega_p$) and ENZ wavelength after excitations, which are determined from Lorentz--Drude fitting of the integrated transient absorption spectra over time. (c): Temporal evolution of the extracted P2 polaron absorption traces (scattered points) from the Lorentz--Drude fitting, which is well fitted with the rate equations, as indicated by the solid line. }
\end{center}
\end{figure*}

Furthermore, a set of rate equations is proposed to reveal ultrafast dynamics of polaron excitation considering the physics above, as expressed in \textbf{Eqns.} (4-6), which can be referred to the modified two--temperature model \cite{Jia2022}. In more detail, Eqn. (4) expresses the generation of photoexcited electrons ($G(t)$) from the laser irradiation and their disappearance due to the polaron formation and electron--lattice scattering interaction, where $G(t)=\frac{2}{s\sqrt{2\pi}}\exp(-2t^2/s^2)$ with the laser pulse duration $s$. Eqn. (5) describes the temporal behavior of phonon population, which is generated by electron--lattice scattering interaction and disappears by polaron formation \cite{Carneiro2017}. Eqn. (6) describes the formation and decoupling of the polaron states.
\begin{align}
\frac{dn_e}{dt} = G(t) -\frac{n_e - n_p}{\tau_{ep}}-\frac{n_e n_p}{\tau_{pol}}, \\
\frac{dn_{p}}{dt} = \frac{n_e - n_{p}}{\tau_{ep}}-\frac{n_e n_{p}}{\tau_{pol}}, \\
\frac{dn_{pol}}{dt} = \frac{n_e n_p}{\tau_{pol}}-\frac{n_{pol}}{\tau_{dcp}},	
\end{align}

where $n_{e}$, $n_{p}$, and $n_{pol}$ are the populations of excited electrons, phonons, and polarons, respectively, and $\tau_{ep}$, $\tau_{pol}$, $\tau_{depl}$ are the time constant of electron--phonon coupling constant, the time constant of polaron formation, and the time constant of polaron decoupling, respectively. The decay of excited electrons due to the electron--phonon scattering is described as $(n_{e}-n_{p})/\tau_{ep}$. The polaron formation is allowed to occur as a decrease in population of both excited states of electrons and phonons, and is expressed as $(n_{e}n_{p})/\tau_{pol}$. The polaron decoupling is described by $n_{pol}/\tau_{dcp}$. 

In order to solve the rate equations above, the initial value of the electron population should be provided. Assume that each photon effectively excites an electron to couple with the phonon, thereby forming a polaron. The electron density $n_e$ can be estimated as $n_e$ $\sim$ $F$/$hvd$ [1-exp(-$\beta d$)], where $F$ is the laser fluence, $hv$ is the incident photon energy, and $d$ is the sample thickness, $\beta$ is the absorption coefficient \cite{Jia2022}. Our calculations show that $n_e$=6.67 $\times$ 10$^{19}$ cm$^{-3}$ at $F$=2.12 mJ/cm$^2$. Then, three time constants, $\tau_{ep}$, $\tau_{pol}$, and $\tau_{dcp}$, can be determined by fitting the transient polaron absorption traces in Fig. 4(c).


Figure 4(c) shows that the extracted polaron absorption trace at $\sim$1081 nm can be well reproduced by fitting to the rate equations above, as indicated by the solid line. The fitting yields $\tau_{ep}$ values of 40 fs and $\tau_{pol}$ values of 80 fs consistent with previous reports \cite{An2004, Ge1998, Wang2015, Moses2000}. Moreover, the time constant of polaron decoupling is estimated to be 278 fs. These results indicate that the ENZ wavelength shift in the PEDOT films can be modulated via polaron excitation within the sub--picosecond time scale.

\section{Conclusions}
In summary, we have demonstrated the feasibility of dynamic tuning of the ENZ wavelength in $p$--type PEDOT films via polaron excitation. An ENZ wavelength shift of $\sim$150 nm was achieved in the EG--modified PEDOT film under laser irradiation of 2.12 mJ/cm$^2$. Such a shift is attributed to a transient increase of hole density due to polaron formation following optical excitation. Moreover, ultrafast dynamics of polaron excitation using femtosecond time--resolved pump--probe spectroscopy reveals a time constant of 80 fs for polaron buildup and a time constant of 278 fs for the polaron decoupling. These findings suggest that an ENZ wavelength shift can be modulated within the sub--picosecond range, enabling potential applications in ultrafast optical signal processing, ENZ photonics with frequency shifts, and ENZ--based multiband communication devices.

\section{METHODS}
\textbf{Sample fabrication}: The PEDOT:PSS (Clevios PH1000) (PEDOT) was spin--cast onto 1.5$\times$1.5 cm$^2$ glass substrates for three times to fabricate the PEDOT film. For ethylene glycol (EG) modified PEDOT film, EG with a 5\% volume fraction was added to the PEDOT:PSS solution to remove PSS for preparing the EG--modified PEDOT solution, which was then spin-cast in the same manner as the PEDOT film. Note that PSS acts as both a soluble template and a charge-balancing counterion for PEDOT.


\textbf{Materials characterization}: The electrical properties of PEDOT and EG--modified films were measured by Hall measurements (LakeShore 7604) Van der Pauw method, where an gallium--based metal droplets were applied to improve ohmic contact between the film and probe. Reflectance and transmittance spectra were acquired using a spectrophotometer (Agilent–Cary–7000), and the complex permittivity was extracted from the transmittance spectrum by fitting data with the Drude--Lorentz model. A stylus profiler (Bruker DEKTAK--XT) was used to measure film thickness, and an ultraviolet absorption spectroscopy (Agilent--Cary--7000) was used to detect the information of the valence band. 

\textbf{Transient VIS-NIR absorption spectroscopy}: Transient VIS-NIR absorption spectroscopy was employed to investigate the ultrafast polaron dynamics. The fundamental laser pulse with wavelength at 800 nm was generated by a Ti: sapphire laser (Coherent Inc) with a pulse repetition rate of 1 kHz and a pulse width of 90 fs. This fundamental pulse was divided into two beams: one was used as the pump laser, and the other was directed at a CaF$_2$ and a thick sapphire crystal to produce a white--light continuum of 550--750 nm and 850--1450 nm, used as the probe laser. The pump beam spot diameter was approximately 300 um. The pump intensity was carefully adjusted to avoid damage to the thin film, where a maximum pumping intensity of 2.12 mJ/cm$^2$ was identified as the damage threshold. The instrument response function (IRF) for the system was 100 fs.

\section*{Acknowledgements}
This work was supported by the Major Program of Zhejiang Provincial Natural Science Foundation (LD24E020004).

\section*{Author Contributions Statement}
H. L and J. J proposed the concept. H. L prepared the sample and measured the VIS-NIR spectra of the sample. H. L and M. J performed the transient measurements. H. L, J. J and H. Y analyzed the experimental results. J. J proposed the model. and H. L simulated the data. H. Y are responsible for all these parts. H. L and J. J wrote this manuscript. H. Y and J. J revised this manuscript.

\section{Competing interests}
The authors declare no competing interests.

\begin{thebibliography}{9}
\bibitem{Fomra2024} D. Fomra, A. Ball, S. Saha, J. Wu, Md. Sojib, A. Agrawal, H. J. Lezec, and N. Kinsey. Nonlinear optics at epsilon near zero: From origins to new materials. Appl. Phys. Rev. {\bf11}, 011317, (2024).

\bibitem{Reshef2019} O. Reshef, I. D. Leon, M. Z. Alam, and R. W. Boyd, Nonlinear optical effects in epsilon--near--zero media. Nat. Rev. Mater. {\bf4}, 535--551 (2019). 

\bibitem{Kinsey2019} N. Kinsey, C. DeVault, A. Boltasseva, and V. M. Shalaev, Near--zero--index materials for photonics, Nat. Rev. Mater. {\bf4}, 742–760 (2019).

\bibitem{Zhao2015} H. Zhao, Y. Wang, A. Capretti, L. D. Negro, and J. Klamkin, Broadband electroabsorption modulators design based on epsilon-near-zero indium tin oxide. IEEE Journal of Selected Topics in Quantum Electronics, {\bf21}, 3300207 (2015).

\bibitem{Jia2024} J. Jia, T. Yagi, Y. Shigesato, Thermal conduction in polycrystalline or amorphous transparent conductive oxide films. Sol. Energy Mater. Sol. Cells {\bf271}, 112872 (2024). 

\bibitem{Kinsey2015} N. Kinsey, C. DeVault, J. Kim, M. Ferrera, V. M. Shalaev, and A. Boltasseva, Epsilon-nearzero Al-doped ZnO for ultrafast switching at telecom wavelengths. Optica {\bf2}, 616–622, 2015.

\bibitem{Guo2017} Q. Guo, Y. Cui, Y. Yao, Y. Ye, Y. Yang, X. Liu, S. Zhang, X. Liu, J. Qiu, and H. Hosono. A solution-processed ultrafast optical switch based on a nanostructured epsilon near-zero medium. Adv. Mater. {\bf29}, 1700754 (2017).

\bibitem{Jia2021} W. Jia, M. Liu, Y. Lu, X. Feng, Q. Wang, X. Zhang, Y. Ni, F. Hu, M. Gong, X. Xu, Y. Huang, W. Zhang, Y. Yang, and J. Han, Broadband terahertz wave generation from an epsilon--near--zero material. Light Sci. Appl. {\bf10}, 11 (2021).

\bibitem{Alam2016} M. Z. Alam, I. D. Leon, and R. W. Boyd, Large optical nonlinearity of indium tin oxide in its epsilon-near-zero region. Science, {\bf352}, 795–797 (2016).

\bibitem{Campione2013} S. Campione, D. de Ceglia, M. A. Vincenti, M. Scalora, and F. Capolino, Electric field enhancement in $\epsilon$-near-zero slabs under TM-polarized oblique incidence, Phys. Rev. B {\bf87}, 035120 (2013).

\bibitem{Capretti2015} A. Capretti, Y. Wang, N. Engheta, and L. D. Negro. Comparative study of second-harmonic generation from epsilon--nearzero indium tin oxide and titanium nitride nanolayers excited in the near--infrared spectral range. ACS Photonics {\bf2}, 1584–1591, 2015.

\bibitem{Yang2019} Y. Yang, J. Lu, A. Manjavacas, T. S. Luk, H. Liu, K. Kelley, J.-P. Maria, E. L. Runnerstrom, M. B. Sinclair, S. Ghimire, and I. Brener. High-harmonic generation from an epsilon-near-zero material. Nat. Phys. {\bf15}, 1022 (2019).
 
\bibitem{Fatti2000} N. D. Fatti, C. Voisin, M. Achermann, S. Tzortzakis, D. Christofilos, and F. Vallée, Nonequilibrium electron dynamics in noble metals. Phys. Rev. B {\bf61}, 16956–16966 (2000).

\bibitem{Caspani2016} L. Caspani, R. P. Kaipurath, M. Clerici, M. Ferrera, T. Roger, J. Kim, N. Kinsey, M. Pietrzyk, A. Di Falco, V. M. Shalaev, A. Boltasseva, and D. Faccio, Enhanced nonlinear refractive index in epsilon-near-zero materials. Phys. Rev. Lett. {\bf116}, 233901 (2016).

\bibitem{Zhou2020} Y. Zhou, M. Z. Alam, M. Karimi, J. Upham, O. Reshef, A. E. Willner, and R. W. Boyd, Broadband frequency translation through time refraction in an epsilon-near-zero material, Nat. Commun. {\bf11}, 2180 (2020).

\bibitem{Yang2017} Y. Yang, K. Kelley, E. Sachet, S. Campione, T. S. Luk, J. -P. Maria, M. B. Sinclair, and I. Brener, Femtosecond optical polarization switching using a cadmium oxide-based perfect absorber. Nat. Photonics {\bf11}, 390, 2017.

\bibitem{Wang2019} H. Wang, K. Du, C. Jiang, Z. Yang, L. Ren, W. Zhang, S. J. Chua, and T. Mei, Extended drude model for intraband-transition-induced optical nonlinearity. Phys. Rev. Applied {\bf11}, 064062 (2019).   


\bibitem{LiuC2021} C. Liu, M. Z. Alam, K. Pang, K. Manukyan, J. R. Hendrickson, E. M. Smith, Y. Zhou, O. Reshef, H. Song, R. Zhang, H. Song, F. Alishahi, A. Fallahpour, A. Almaiman, R. W. Boyd, M. Tur, and A. E. Willner, Tunable Doppler shift using a time-varying epsilon-near-zero thin film near 1550 nm. Optics Letters {\bf46}, 3444 (2021). 

\bibitem{Bruno2020} V. Bruno, S. Vezzoli, C. DeVault, E. Carnemolla, M. Ferrera, A. Boltasseva, V. M. Shalaev, D. Faccio, and M. Clerici. Broad frequency shift of parametric processes in epsilon-near-zero time-varying media, Applied Sciences {\bf10}, 1318 (2020).

\bibitem{Bohn2021} J. Bohn, T. S. Luk, S. Horsley, and E. Hendry, Spatiotemporal refraction of light in an epsilon-near-zero indium tin oxide layer: frequency shifting effects arising from interfaces. Optica {\bf8}, 1532 (2021).

\bibitem{Hu2023} Q. Hu, X. Yu, H. Liu, J. Qiu, W. Tang, S. Liang, L. Li, M. Du, J. Jia, H. Ye, Tunable organic ENZ materials with large optical nonlinearity. ACS Photonics {\bf10}, 3612 (2023).    

\bibitem{Ahsin2024} A. Ahsin, I. Ejaz, S. Sarfaraz, K. Ayub, H. Ma, Polaron formation in conducting polymers: A novel approach to designing materials with a larger NLO response. ACS Omega {\bf9}, 14043 (2024). 

\bibitem{Sun2015} K. Sun, S. Zhang, P. Li, Y. Xia, X. Zhang, D. Du, F. H. Isikgor, and J. Ouyang, Review on application of pedots and pedot: Pss in energy conversion and storage devices. J. Mater. Sci.: Mater. Electron. {\bf26}, 4438–4462 (2015).

\bibitem{Bredas1982} J. L. Brédas, R. R. Chance, and R. Silbey, Comparative theoretical study of the doping of conjugated polymers: Polarons in polyacetylene and polyparaphenylene. Phys. Rev. B, {\bf26}, 5843–-5854, 1982.

\bibitem{Wang2018} X. Wang, X. Zhang, L. Sun, D. Lee, S. Lee, M. Wang, J. Zhao, Y. Shao-Horn, M. Dinca, T. Palacios, and K. K. Gleason, High electrical conductivity and carrier mobility in oCVD pedot thin films by engineered crystallization and acid treatment. Sci. Adv. {\bf4}, eaat5780 (2018).

\bibitem{Kim2012} N. Kim, B. H. Lee, D. Choi, G. Kim, H. Kim, J. -R. Kim, J. Lee, Y. H. Kahng, and K. Lee, Role of interchain coupling in the metallic state of conducting polymers. Phys. Rev. Lett. {\bf109}, 106405 (2012).

\bibitem{Zozoulenko2019} I. Zozoulenko, A. Singh, S. K. Singh, V. Gueskine, X. Crispin, and M. Berggren, Polarons, Bipolarons, And Absorption Spectroscopy of PEDOT. ACS Appl. Polym. Mater. {\bf1}, 83 (2019). 

\bibitem{Karki2022} A. Karki, G. Cincotti, S. Chen, V. Stanishev, V. Darakchieva, C. Wang, M. Fahlman, and M. P. Jonsson, Electrical tuning of plasmonic conducting polymer nanoantennas. Adv. Mater. {\bf34},  2107172 (2022).

\bibitem{Chen2020} S. Chen, E. S. H. Kang, M. S. Chaharsoughi, V. Stanishev, P. Kühne, H. Sun, C. Wang, M. Fahlman, S. Fabiano, V. Darakchieva, M. P. Jonsson, Conductive polymer nanoantennas for dynamic organic plasmonics. Nat. Nanotechnol. {\bf15}, 35 (2020).


\bibitem{Kim2008} J. Kim, S. Park, and N. F. Scherer, Ultrafast dynamics of polarons in conductive polyaniline: Comparison of primary and secondary doped forms. J. Phys. Chem. B {\bf112}, 15576–15587 (2008).

\bibitem{Kalagi2016} S. S. Kalagi, P. S. Patil. Secondary electrochemical doping level effects on polaron and bipolaron bands evolution and interband transition energy from absorbance spectra of PEDOT: PSS thin films. Synth. Met. {\bf220}, 661–666 (2016).

\bibitem{Bubnova2014} O. Bubnova, Z. U. Khan, H. Wang, S. Braun, D. R. Evans, M. Fabretto, P. Hojati-Talemi, D. Dagnelund, J. -B. Arlin, Y. H. Geerts, S. Desbief, D. W. Breiby, J. W. Andreasen, R. Lazzaroni, W. M. Chen, I. Zozoulenko, M. Fahlman, P. J. Murphy, M. Berggren, and X. Crispin, Semi--metallic polymers, Nat. Mater. {\bf13}, 190 (2014).


\bibitem{Lee1993} K. H. Lee, A. J. Heeger, and Y. Cao, Reflectance of polyaniline protonated with camphor sulfonic-acid disordered metal on the metal--insulator boundary. Phys. Rev. B {\bf48}, 14884--14891, (1993).

\bibitem{Chen2019} S. Chen, P. Kuhne, V. Stanishev, S. Knight, R. Brooke, I. Petsagkourakis, X. Crispin, M. Schubert, V. Darakchieva, and M. P. Jonsson. On the anomalous optical conductivity dispersion of electrically conducting polymers: ultra--wide spectral range ellipsometry combined with a Drude--Lorentz model. J. Mater. Chem. C, {\bf7}, 4350–-4362 (2019).


\bibitem{Emin1993} D. Emin, Optical properties of large and small polarons and bipolarons. Phys. Rev. B, {\bf48}, 13691–13702, 1993.

\bibitem{Jia2022} J. Jia, T. Yagi, M. Mizutani, N. Yamada, and T. Makimoto. Revealing the simultaneous increase in transient transmission and reflectivity in InN. J. Appl. Phys. {\bf132}, 165702 (2022).

\bibitem{Carneiro2017} L. M. Carneiro, S. K. Cushing, C. Liu, Y. Su, P. Yang, A. P. Alivisatos, and S. R. Leone, Excitation wavelength--dependent small polaron trapping of photoexcited carriers in $\alpha$--Fe$_2$O$_3$. Nat. Mater. {\bf16}, 819, 2017.

\bibitem{An2004} Z. An, C. Q. Wu, and X. Sun, Dynamics of photogenerated polarons in conjugated polymers, Phys. Rev. Lett. {\bf93}, 216407 (2004).

\bibitem{Ge1998} N. H. Ge, C. M. Wong, R. L. Lingle, J. D. McNeill, K. J. Gaffney, and C. B. Harris. Femtosecond dynamics of electron localization at interfaces. Science {\bf279}, 202--205, 1998.
 
\bibitem{Wang2015} T. Wang, C. Caraiani, G. W. Burg, and W. -L. Chan, From two--dimensional electron gas to localized charge: Dynamics of polaron formation in organic semiconductors. Phys. Rev. B {\bf91}, 041201(R) (2015).

\bibitem{Moses2000} D. Moses, A. Dogariu, and A. J. Heeger. Ultrafast photoinduced charge generation in conjugated polymers. Chem. Phys. Lett. {\bf316}, 356--360, (2000).

\bibitem{Liu2021} Z. Liu, L. Wu, J. Qian, J. Peng, R. Liu, Y. Xu, X. Shi, C. Qi, and S. Ye. Tuned transport behavior of the IPA--treated PEDOT:PSS flexible temperature sensor via screen printing. J. Electron. Mater. {\bf50}, 2356–-2364, (2021).
\end{thebibliography}
\end{document}